\documentclass[prb,twocolumn,superscriptaddress,showpacs,preprintnumbers,amsmath,amssymb,floatfix,a4paper]{revtex4-1}
\usepackage{graphicx}           
\usepackage{dcolumn}     
\usepackage{bm}                     
\usepackage{amsmath}
\usepackage{amssymb}
\usepackage{amsfonts}
\usepackage{csquotes}
\usepackage{xcolor}
\begin{document}
\title{Dynamics of hydrogen guests in ice XVII nanopores }
\author{Leonardo del Rosso}
\affiliation{ISC--CNR, via Madonna del Piano 10, I-50019 Sesto Fiorentino, Italy}
\author{Milva Celli}
\altaffiliation{present affiliation: Istituto di Fisica Applicata \enquote{Nello Carrara}, IFAC--CNR} 
\affiliation{ISC--CNR, via Madonna del Piano 10, I-50019 Sesto Fiorentino, Italy}
\author{Daniele Colognesi}
\altaffiliation{present affiliation: Istituto di Fisica Applicata \enquote{Nello Carrara}, IFAC--CNR} 
\affiliation{ISC--CNR, via Madonna del Piano 10, I-50019 Sesto Fiorentino, Italy}
\author{Svemir Rudi\' c}
\affiliation{ISIS Facility, STFC Rutherford Appleton Laboratory, Chilton, Oxfordshire OX11 0QX, United Kingdom}
\author{Niall J. English}
\affiliation{School of Chemical and Bioprocess Engineering, University College Dublin, Belfield, Dublin 4, Ireland}
\author{Christian J. Burnham}
\affiliation{School of Chemical and Bioprocess Engineering, University College Dublin, Belfield, Dublin 4, Ireland}
\author{Lorenzo Ulivi}
\altaffiliation{present affiliation: Istituto di Fisica Applicata \enquote{Nello Carrara}, IFAC--CNR} 
\altaffiliation{and Michigan Technological University, Houghton, MI, (USA)} 
\email{e-mail: l.ulivi@ifac.cnr.it}
\affiliation{ISC--CNR, via Madonna del Piano 10, I-50019 Sesto Fiorentino, Italy}

\date{\today}
%
%
%
%
%
\begin{abstract}
The present high-resolution inelastic neutron scattering experiment on ice XVII, containing molecular hydrogen with 
different ortho/para ratio, allows to assign the H$_2$ motion spectral bands to rotational and center-of-mass translational 
transitions of either {\it para}- or {\it ortho}-H$_2$.
Due to its structure, ice XVII confines H$_2$ molecules to move in spiral channels of molecular size.
Reported data demonstrate that H$_2$ molecules rotate almost freely in these nanometric channels, though showing larger perturbation 
than in clathrate hydrates, and perform a translational motion exhibiting two low frequency excitations.
The agreement between experimental spectra and corresponding molecular dynamics results clearly enables to portray
a picture of the confined motions of a hydrophobic guest within a metastable ice framework, i.e. ice XVII.
\end{abstract}
%
%
%
\maketitle
%
%
%
%
%
Ice XVII is a newly discovered form of solid water obtained from the so-called $C_0$-phase of the H$_2$-H$_2$O binary compound, 
quenched at a temperature $T=77$ K and ambient pressure, after letting the hydrogen molecules diffuse out of the crystal \cite{delRosso16}.
It is a pure water crystal, metastable at ambient pressure if maintained below 130 K.
Its crystal structure is intrinsically porous and presents accessible channels where hydrogen molecules have been located during the 
production and where other molecules (belonging to hydrogen  or another gas) can be absorbed again, confined in an essentially one dimensional 
geometry \cite{delRosso16a}.
The diameter of these channels, measured from the center of the oxygen atoms, is about 6.10 \AA.
This exotic and low-density water crystal adds to the list of solid structures of water possibly stable at negative pressure \cite{Huang16}.

The study of the dynamics of hydrogen molecules in nano-confinement, that is intrinsically quantum-mechanical, is of great importance from both a practical and fundamental point of view.
In hydrogen clathrate hydrates, which possess a similar chemical composition, 
although a different structure and stoichiometry, the hydrogen molecules confined in nearly spherical cages \cite{Sloan97} 
perform an almost free rotation and a deeply non-harmonic center-of-mass (CoM) vibrational motion (rattling), both of which have been experimentally investigated by inelastic neutron scattering \cite{Ulivi07,Xu11a,Xu13a,Celli13,Colognesi13} and Raman scattering \cite{Giannasi08a,Strobel09,Giannasi11,Zaghloul12}.
In this paper we discuss the results of a combined experimental and simulation study on the dynamics of the H$_2$ guests in D$_2$O ice XVII and of the D$_2$O host lattice.

Samples  for the present Inelastic Neutron Scattering (INS) experiment were produced at ISC--CNR using D$_2$O and H$_2$, as described in Ref.~\onlinecite{delRosso16}.
This isotopic composition is chosen in order to exploit the intrinsic advantage of the large incoherent neutron scattering cross-section of the proton (compared to both D and O), thus allowing for a relatively simple access to the self dynamics of molecular hydrogen.
Measurements have been performed at $T=15$ K on TOSCA, a high energy resolution ($ \Delta E $) spectrometer at ISIS (UK) characterized by $1.5 \% \lesssim  \Delta E/ E_i \lesssim 2.5 \%$, with $E_i$ being the incoming neutron energy.
Raw time-of-flight data have been transformed into energy-transfer spectra, taking into account both the correction for the kinematic factor and the normalization for the incoming neutron flux.

We have measured the spectra for three different gas-charged samples and one reference sample of pure deuterated ice XVII.
Initially, we probed the material as prepared, i.e. the metastable H$_2$-D$_2$O compound in the $C_0$ phase, quenched at low temperature and ambient pressure.
This material, as for structure and composition, does not differ much from ice XVII when refilled with H$_2$, apart for small  possible nitrogen impurities \cite{delRosso16a} which do not give visible signal in this experiment.
After a thermal treatment at about 120 K that removes all the guests molecules (described in Ref.~\onlinecite{delRosso16}), we have recorded a spectrum of pure deuterated ice XVII at $T= 15$ K.
Subsequently, we have measured two spectra of ice XVII loaded, respectively, with normal ({\it n}-H$_2$) and {\it para}-enriched hydrogen ({\it p}-rich H$_2$).
Due to the limited working pressure of our aluminum sample cell, a quite low temperature (i.e. $T= 20$ K) has been chosen for the gas loading 
processes.
According to previous work \cite{delRosso16,delRosso17}, this produces ice XVII with a H$_2$/D$_2$O molar ratio of about 25\%, that is, half filling of the H$_2$ crystallographic sites.

\begin{figure}[b!]
\includegraphics[viewport= 0.5cm 2cm 20cm 27.5cm, width=0.9\columnwidth]{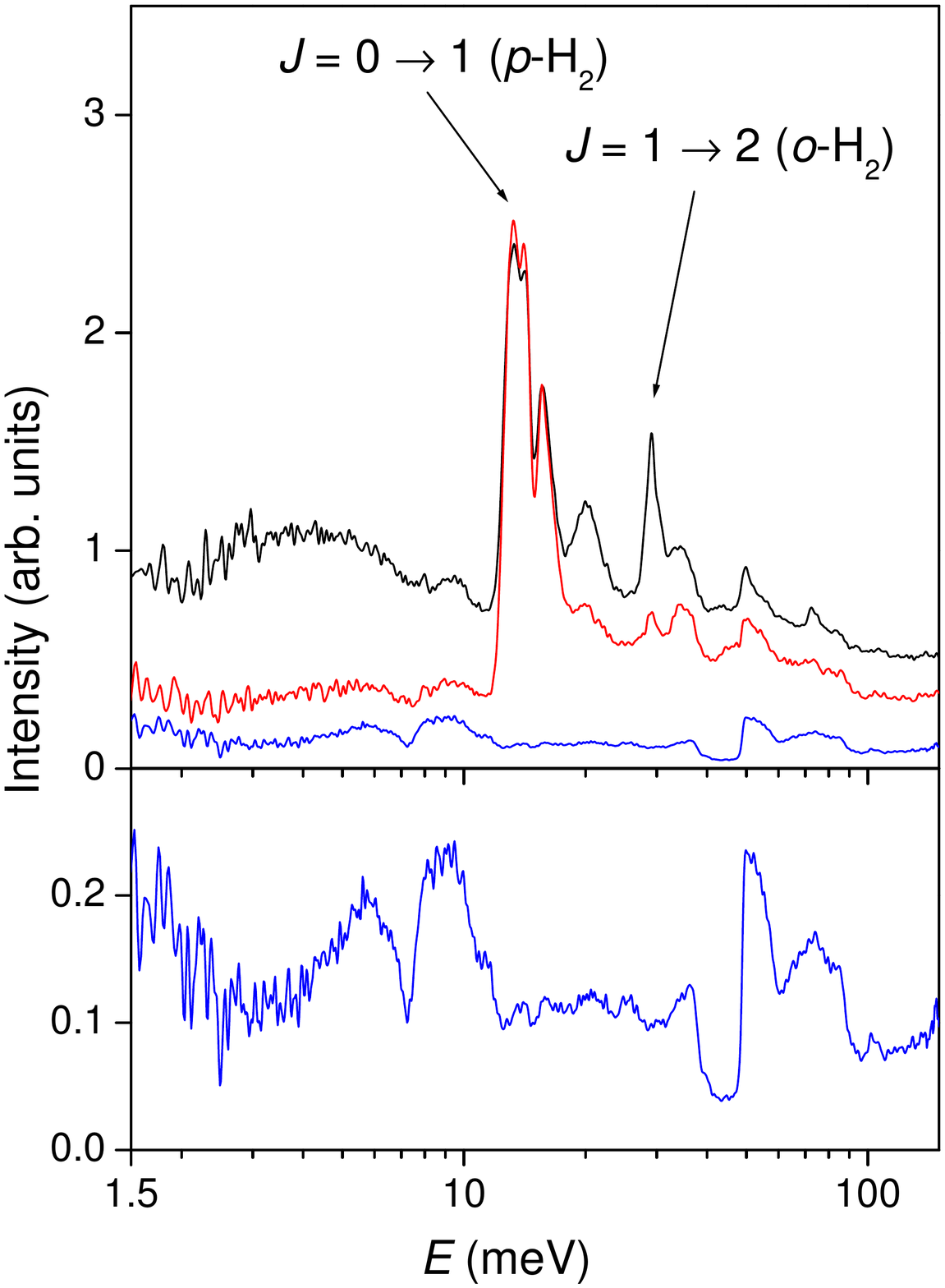}
\caption{ (Color online) Bottom panel: INS spectrum of deuterated ice XVII. 
Top panel:  the two upper traces are the spectra of ice XVII refilled with normal (black line)  
and para-enriched (red line) hydrogen.
Blue line represents the same spectrum as in the bottom panel replotted here for an easier comparison.
All spectra have been recorded at $T=15$ K.
The logarithmic horizontal scale implies that the width of the bands in different spectral regions are not readily comparable by eye.  
\label{f.RawSpectra} }
\end{figure}

Measured spectra, before any analysis, are shown in Fig.~\ref{f.RawSpectra} on a logarithmic horizontal axis.
%
%
Many features due to translational and librational excitations of the lattice are already recognizable in the spectrum of empty ice XVII (blue line).
The other two spectra from the hydrogen-charged samples (black {\it n}-H$_2$, red {\it p}-rich H$_2$), present, in addition, several narrow and intense bands due to the molecular hydrogen dynamics.
The two main H$_2$-molecule rotational features are readily identified in the spectrum and assigned, namely  the strong $ J \! = \! 0 \! \rightarrow \! 1 $ rotational band of  {\it para}-hydrogen  ({\it p}-H$_2$), i.e. a triplet at $\simeq$ 14 meV, and the $ J \! = \! 1 \! \rightarrow \! 2 $ rotational band of {\it ortho}-hydrogen  ({\it o}-H$_2$) around 29 meV.

For the identification of the CoM excitations a subtraction procedure is applied, similar to that described in Ref.~\onlinecite{Ulivi07}.
First, the hydrogen contribution was obtained subtracting the weak reference spectrum (i.e. that of empty ice XVII plus aluminum cell, adjusted by considering self-shielding attenuation).
The resulting  H$_2$ spectrum can be modeled considering the neutron scattering cross section for the two hydrogen-molecule spin isomers, and its dependence on the rotational transitions \cite{Lovesey84}.
After neglecting the coherent part of the neutron scattering, the double differential scattering cross-section becomes proportional to the self part of the dynamical structure factor for the CoM motion, $S_{\mathrm{self}} (Q,\omega)$, and can be written as \cite{Colognesi04}:
\begin{equation}
\frac{d^2 \sigma}{d\Omega d\omega}
=\frac{ k_\mathrm{f}}{ k_\mathrm{i}}S_{\mathrm{self}} (Q,\omega) 
\otimes \sum_{JJ'}\delta(\omega-\omega_{JJ'})\nu(J,J',Q),
\label{e.theonlyone}
\end{equation}
where $k_\mathrm{i}$ and $k_\mathrm{f}$ stand for the initial and final neutron wavevectors, respectively, $\omega$ is simply $\hbar^{-1} E$, and the symbol $\otimes$ represents a convolution product.
This expression holds whenever the hypothesis of decoupling between rotational and CoM motions is satisfied.
The Dirac delta functions reported in the previous equation are centered at the rotational transition energies, $\hbar \omega_{JJ'}$, while $\nu(J,J',Q)$ (called molecular form factor) depends on both the momentum transfer $Q$ and the rotational transition $J \! \rightarrow \! J'$ undergone by molecule.
Therefore the expected neutron spectrum is made of a comb of CoM excitations, replicated and shifted by the energy of any possible rotational excitation of the single molecule.
The molecular form factors $\nu(J,J',Q)$ can be easily calculated assuming either a rigid rotor \cite{Young64} or a rotating harmonic oscillator \cite{Zoppi93}  model.

\begin{figure}[h!]
\includegraphics[bb= 0.5cm 2cm 20cm 27.5cm, width=0.9\columnwidth]{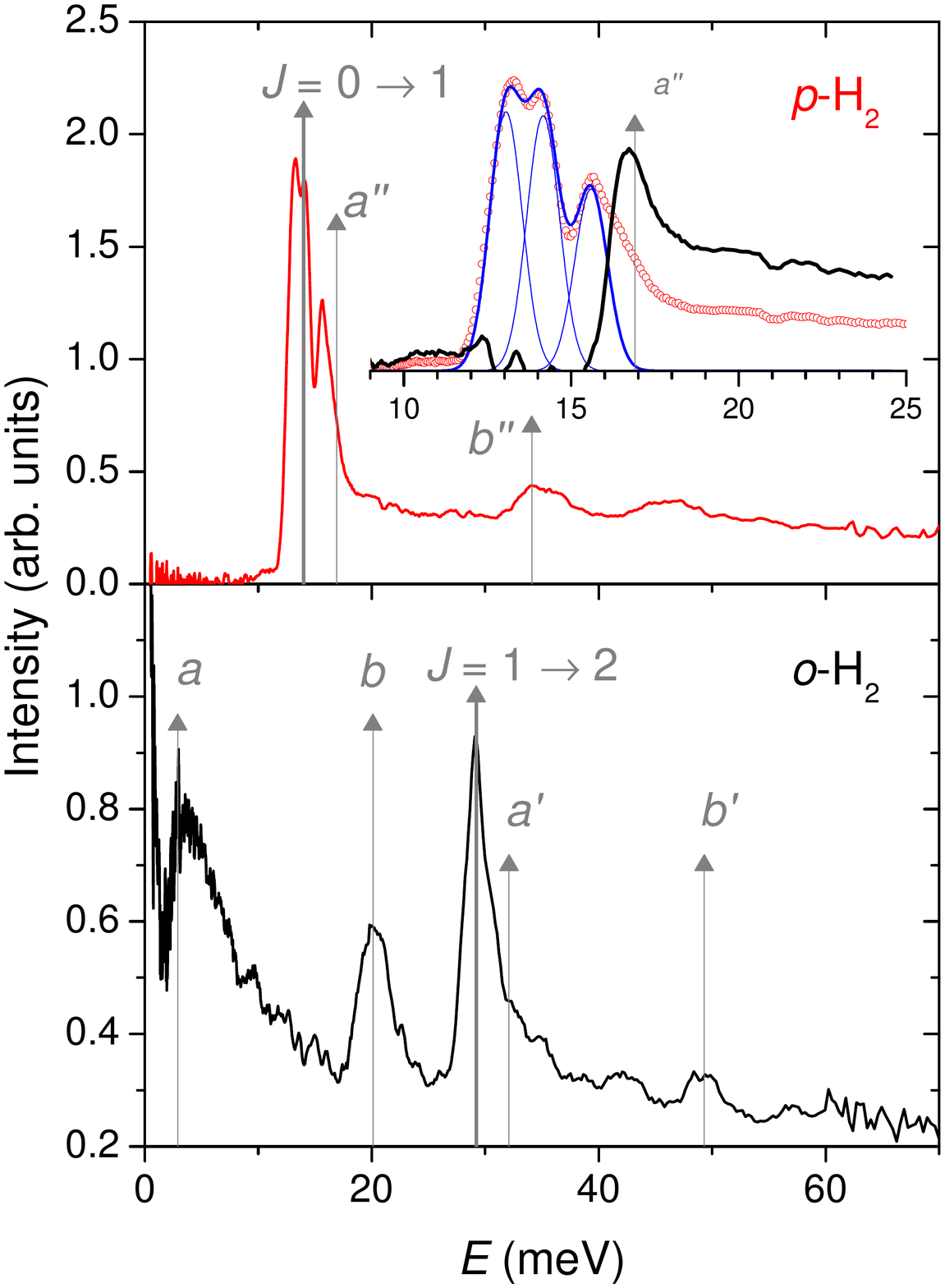}
\caption{(Color online)  Spectra of pure  {\it p}-H$_2$ (top panel) and pure  {\it o}-H$_2$ (bottom panel).
The two observed CoM excitations are marked with $a$ and $b$, while we use  $a^{\prime}$,  $b^{\prime}$,  
and
$a^{\prime\prime}$, 
$b^{\prime\prime}$  for the combinations of these excitations with $J \! = \! 1 \! \rightarrow \! 2$ 
and $J \! = \! 0 \! \rightarrow \! 1$ rotation, respectively.
In the inset we show the decomposition of the  $J \! = \! 0 \! \rightarrow \! 1$ rotational lines into three components.
The residuals (scaled by a factor two and plotted as a black line in the inset) highlight the presence of the intensity due to 
the $a^{\prime\prime}$ mode.  
\label{f.oH2_pH2_Spectra} }
\end{figure}

Molecular hydrogen trapped in ice XVII channels is a non-equilibrium mixture of {\it o}-H$_2$ and {\it p}-H$_2$, in a concentration which is essentially invariant in time, being the conversion rate extremely low.
At the low temperature values typical of the present experiment, only the lowest rotational state for each species is populated (namely, $J\!=\!0$ for {\it p}-H$_2$ and  $J\!=\!1$ for {\it o}-H$_2$) and so few transitions contribute to the spectrum in the frequency region of interest, namely the rotationally-elastic $J \! = \! 1 \! \rightarrow \! 1$ and the inelastic $J \! = \! 1 \!  \rightarrow \! 2$ transition of {\it o}-H$_2$, plus the inelastic transition $ J \!=\! 0 \!\rightarrow \! 1 $ of {\it p}-H$_2$. 
We remark that the $J \! = \! 0 \! \rightarrow  \! 0 $ transition of the {\it p}-H$_2$ molecule, being weighted by the mere coherent cross section, does not contribute appreciably to the observed spectrum.
As a consequence, we assume that the spectral intensity recorded in the energy range below 10 meV is  solely due to ortho molecules.
Being the strong $ J \!=\! 0 \!\rightarrow \! 1 $ band around 14 meV due only to {\it p}-H$_2$, we can extract, by a linear combination of the two measured spectra (samples with {\it n}-H$_2$ and {\it para}-rich H$_2$),  the spectra of pure {\it p}-H$_2$ and {\it o}-H$_2$, represented in Fig.~\ref{f.oH2_pH2_Spectra}.
%
%

This decomposition enables to highlight two intense bands due to the translational excitations of the molecule CoM, which appear at low energy in the spectrum of {\it o}-H$_2$, as combinations with the rotationally elastic $J \! = \! 1 \! \rightarrow \! 1$ line.
These bands, marked with letters $a$ and $b$ in the figure, are quite broad (about 4.0 and 3.5 meV respectively). Band $a$ is asymmetric, extending from 2.2 to 6.2 meV, while $b$ has a more symmetrical shape.
The position of these band is estimated at $2.9(8)$ and $ 20.1(6)$ meV.
According to Eq.~(\ref{e.theonlyone}), the same CoM excitations should give rise to similar combinations with the other rotational lines, shifted by an equal amount. 
This is indeed the case, as can be seen by observing Fig.~\ref{f.oH2_pH2_Spectra}, where the combination bands are evident exactly at the positions where they are expected,  marked $a^\prime$ $b^\prime$  for {\it o}-H$_2$, (bottom panel) and $a^{\prime\prime}$ $b^{\prime\prime}$ for {\it p}-H$_2$ (top panel).

In the inset of Fig.~\ref{f.oH2_pH2_Spectra} we show the fitting of the $J \! = \! 0 \! \rightarrow \! 1$ rotational line, exhibiting three separated components.
The splitting of the rotational band  into a triplet, commonly observed for hydrogen molecules in a confined geometry, is due to the lifting of the threefold degeneracy of the $J \! = \! 1$ rotational level, as consequence of the potential energy anisotropy with respect to the orientation of the H$_2$ molecule.
This rotational triplet is nicely fitted by the sum of three Gaussians, whose energy positions are 13.05, 14.16, and 15.59 meV.
Comparing these values with those measured for the same  transition of  H$_2$ in clathrate hydrates \cite{Ulivi07} (namely, 13.64, 14.44, and 15.14 meV), we observe here a larger splitting (i.e. 2.54 meV instead of 1.50 meV)  proving a stronger potential energy anisotropy.

This fitting procedure also highlights the extra intensity due to the $a^{\prime\prime}$ band. 
Concerning the CoM translational motion of the hydrogen molecule inside the channel, even though this has a spiral shape \cite{delRosso16a} it can be pictured as locally cylindrical.
Therefore two vibrational frequencies are expected, the lower corresponding to the vibration along the cylinder axis, say $z$, and the higher to the $xy$ degenerate mode.
This is the way we have assigned the  $a$ and $b$ bands, observed in the spectra.
In order to verify this assignment, we have performed classical molecular dynamics (MD) and calculated the spectrum of the CoM motion.

MD computation was performed at $T$=50 K, with water potential model TIP4P-2005 \cite{English03} and Alavi's H$_2$ model \cite{Alavi05}, using Partridge-Swenke and Morse intramolecular flexibility for water and H$_2$, respectively.
The average structure of empty ice XVII is known from accurate Rietveld refinement of neutron diffraction data \cite{delRosso16a} and is described by the space group $P6_122$. 
Less certain but at any rate very similar is the structure of the samples containing H$_2$.
A neutron diffraction measurement on the recovered $C_0$ phase \cite{Efimchenko11} determined a lower symmetry space group ($P3_112$). 
On the contrary, Strobel {\it et al.} \cite{Strobel16}, from X-ray diffraction measurements in a diamond anvil cell, recently claimed to have determined that also the high pressure phase $C_0$  exhibits a hexagonal $P6_122$ structure.
The differences between the atomic coordinates of the two structures are numerically minimal, but are substantial to determine the potential energy shape for one H$_2$ molecule hypothetically located in the channel, as clearly discussed in Ref.~\onlinecite{delRosso16a}.
Specifically, the ice lattice configuration based on  the lower symmetry space group $P3_112$ generates three stable positions for the H$_2$ molecule per each hexagonal unit cell. 

For the MD calculation aimed to study the CoM dynamics, we start from an initial configuration made by 384 water molecules, located on the lattice sites of a 
supercell  $4 \times 2 \times 4$ times larger than the conventional orthorhombic unit cell for the space group $P6_122$.
The assumed lattice constants are $a= 6.326$ \AA\ and   $c= 6.080$ \AA\ for the hexagonal cell (corresponding to an orthorhombic cell $6.326\times 10.957 \times 6.080 $  \AA$^3$).
The O atoms of the water molecules in the supercell are placed on their crystallographic positions.
However, the D atoms in deuterated ice XVII are configurationally disordered.
Different random distributions of the framework water deuterons were generated by a computer routine, each consistent with the Bernal-Fowler ice rules \cite{Bernal33} and the periodic boundary conditions outside the supercell.
%
%
\begin{figure}[hbt!]
\includegraphics[bb= 0.5cm 0.5cm 27cm 21cm, width=0.9\columnwidth]{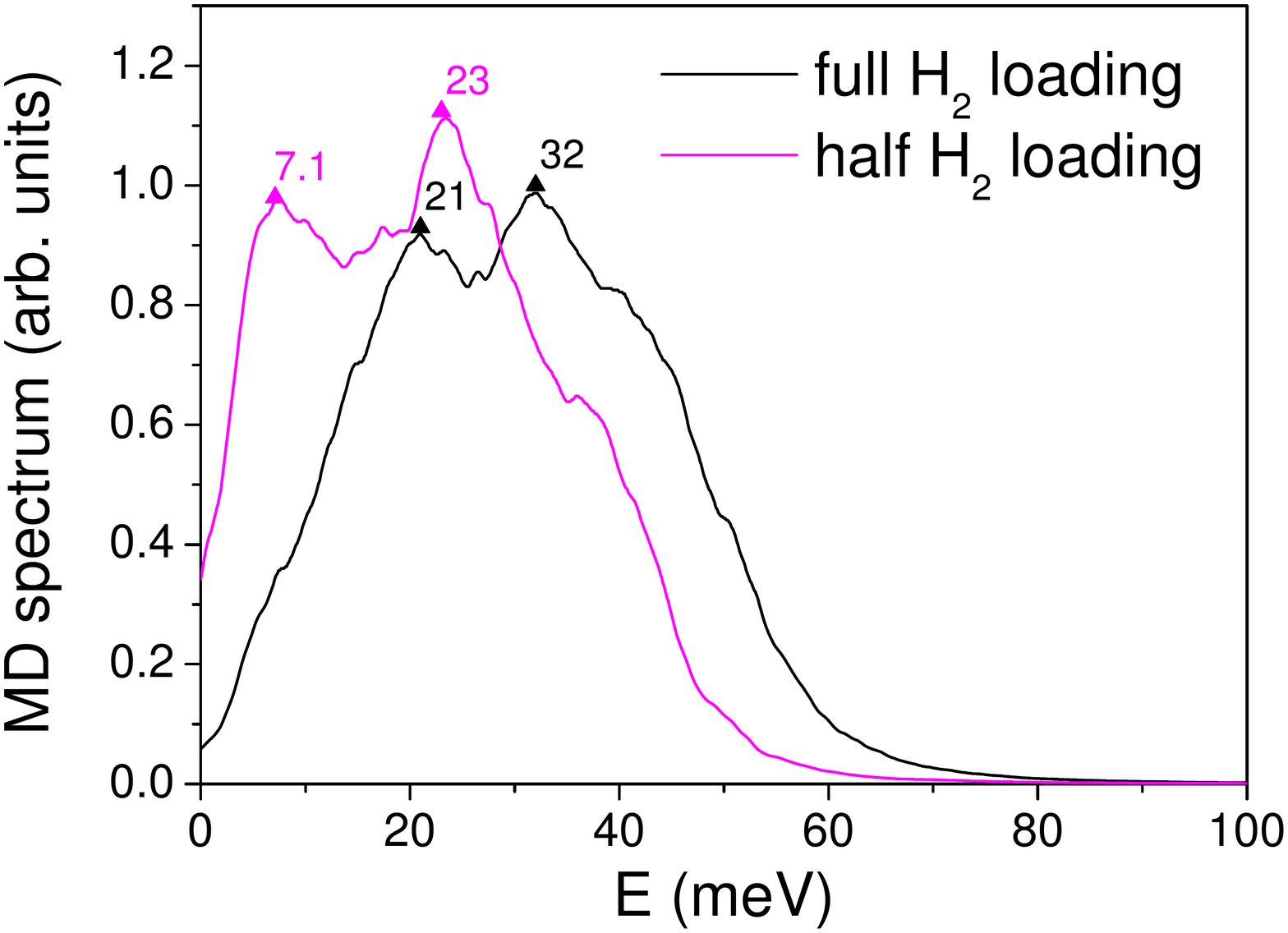} 
\caption{(Color online) 
Spectra of the CoM velocity autocorrelation function, obtained by MD calculation, for completely hydrogen filled  (black line) 
and randomly half-filled (magenta line) ice XVII.
Symbols mark the positions of the maxima of each curve.
\label{f.MDspectra}}
\end{figure}
%
%
A proton configuration with a negligible dipole moment was selected for use in further calculations. 

The spectra calculated for the H$_2$ CoM motion are shown in Fig.~\ref{f.MDspectra}, assuming full hydrogen loading (i.e. 3 H$_2$ molecules for primitive cell) or half loading, (i.e. randomly filling half of the H$_2$ sites).
The comparison with the experimental values shows a semi-quantitative agreement, considering that our samples are in a situation close to half filling.
The disagreement of the low frequency value can be easily imputed to the approximations (classical motion, empirical interaction potentials, structural model etc.) in the computation and, partly, to the large uncertainty of the experimental band energy.
It is interesting to note that, by increasing the loading, the low energy mode essentially disappears and the intensity of a mode at higher energy increases.
This supports the assignment of the modes at about 7 meV and 23 meV to the motion along and across the spiral channels, respectively.
The energy of the mode along the channels is much more sensitive to the amount of loading, which is very reasonable.
\par
\begin{figure}[t!]
\includegraphics[bb= 0.5cm 2.0cm 21cm 27cm, width=\columnwidth]{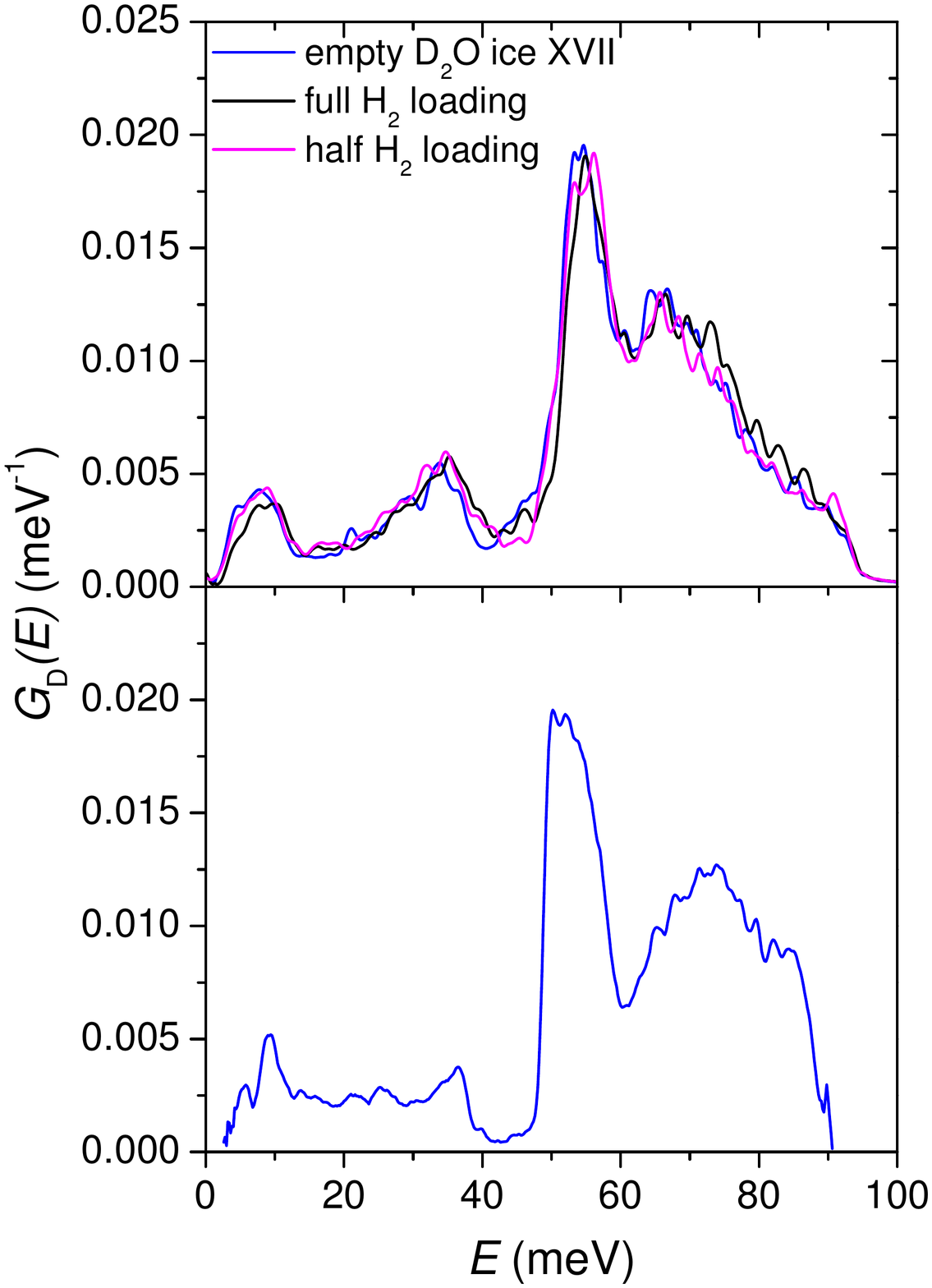}
\caption{(Color online) Experimentally determined (lower panel) and calculated (upper panel) density of phonon states, projected on the deuteron nucleus.
The calculation has been performed for different hydrogen filling.  
Experimental data are derived from the spectra in Fig. \ref{f.RawSpectra} (bottom panel).
\label{f.HPDoS}}
\end{figure}

By means of the same MD computation, we have calculated the D-projected density of phonon states (DoPS), which is compared in Fig.~\ref{f.HPDoS} with the same quantity extracted from our neutron scattering spectra.
The analysis of the experimental data has been done analogously to what presented in Ref.~\onlinecite{Celli12a}.
The first steps are straightforward and consist in the subtraction of the empty container contribution and the correction for self-shielding attenuation, operated via the analytical approach suggested by Agrawal\cite{Agrawal71}.
%
%
%
%
Multiple scattering contaminations made of two inelastic scattering events have been found to be practically negligible in the energy transfer interval of interest (i.e., 2 meV$<E<100$ meV).
Finally, in order to determine and remove the multi-phonon terms in these spectra, we have used a well-known self-consistent procedure\cite{Dawidowski98} already tested with positive experience on a number of systems. 
Such a procedure, performed in the framework of the incoherent approximation, would have been totally justified for a polycrystalline H$_2$O-based material and would have led directly to the proton-projected density of phonon states, due to the much larger cross-section, and lower mass, of the H nucleus than that of the O nucleus. 
For a deuterated sample like ours, however, the rationale of this approximation is based on well-established results obtained on various forms of D$_2$O ice\cite{Li96}. 

Despite the various approximations involved (i.e. a purely incoherent, harmonic, isotropic, and single-site treatment of the multi-phonon terms), the results reveal a satisfactory convergence of the method, allowing for a sound extraction of the one-phonon component of the self-dynamic structure factor, and, finally, for the evaluation of the deuterium-projected density of phonon states in ice XVII, $G_D(E)$, which is reported in Fig.~\ref{f.HPDoS}.
It is worth noting that, in order to obtain an accurate result from a quantitative point of view, the oxygen contribution to the experimental estimate of $G_D(E)$ has been duly evaluated and subtracted.
This contribution has been found to be fully negligible in the librational part of the spectrum, while in the lattice phonon part it turned out to amount to 27.7 \% of the deuterium one. Due to the fact that the lattice-phonon corresponds to a rigid motion of D$_2$O molecules as single units, then the D and O contributions to the density of phonon states in the corresponding spectral zone exhibit exactly the same spectral shape and so the aforementioned corrections to $G_D(E)$ can be simply performed by scaling the intensity of the lattice part of $G_D(E)$ (in the energy transfer range from 0 to 42 meV) by the factor 0.783, obtained by taking into account the different cross section of D$_2$O molecule and D nucleus. 

The calculated DoPS does not depend appreciably on the hydrogen loading, and shows some bands originated by different vibrational modes.
By looking at Fig.~\ref{f.HPDoS}, one can observe that the lattice phonons region (i.e., $E<42$ meV) of $G_D(E)$ 
is different from what can be observed in basically all other ice forms \cite{Li96}, but is quite similar to the lattice band in deuterated sII clathrate hydrate (filled with Ne) \cite{Celli12a}, even though the cleft of the main acoustic peak seems less deep and more asymmetric in ice XVII.
In the librational region (i.e. 45 meV $ <E<90$ meV), one also notes some differences between $G_D(E)$ of ice XVII and that of sII clathrate hydrate.
In particular, the first steep peak, located at about 50 meV, appears more asymmetric and centered at a slightly lower frequency in ice XVII, followed by a dip, at around 60 meV, which looks shallower and less pronounced.
Altogether the comparison with the computed  $G_D(E)$ shows a reasonably good agreement, similar to what obtained for clathrate hydrates of various structures, as discussed in Refs.~\onlinecite{Cimtec} and \onlinecite{Burnham16}.

In this work we have accurately measured the quantum dynamics of a single H$_2$ molecule in the confined geometry of one single nanometric ice XVII channel.
The splitting of the rotational and translational bands is a consequence of the water environment, whose anisotropy appears stronger than in the clathrate hydrate confinement.
As for the CoM translational dynamics, the comparison between the measured spectra and the MD calculations supports the identification of the lowest frequency band as the vibration along the channel direction, while the higher mode corresponds to the motion across the spiral channel. 
Both frequencies are significantly influenced by the hydrogen loading, but this dependence is rather more marked for the vibration along the channel. 
The present data about the guest motion, together with a substantial agreement between the measured and calculated D-projected DoPS of the host molecules, allow to characterize the main features of the vibrational dynamics of this novel inclusion compound. 

\begin{acknowledgments}
We gratefully thank Mr.  Andrea Donati for his skillful technical support in the sample preparation.
We acknowledge the PRIN project ZAPPING,  number 2015HK93L7, granted by the Italian Ministry of Education, Universities and Research (MIUR).  
This work has been performed within the Agreement No.0018318 (02/06/2014) between STFC and CNR,
concerning collaboration in scientific research at the spallation neutron source ISIS
\end{acknowledgments}
%
%
%

\end{document}